# Microstructure and elevated-temperature mechanical properties of refractory AlMo$_{0.5}$NbTa$_{0.5}$TiZr High Entropy Alloy fabricated by powder metallurgy

Yuqiao Li, Junho Lee, Byungchul Kang, Soon Hyung Hong

Department of Materials Science and Engineering

Korea Advanced Institute of Science and Technology, Republic of Korea

**Abstract**

New approaches for the design of alloy systems with multiprincipal elements is recently researched in refractory materials field. However, most research aimed at arc melting process with weakness of coarsening of grains and inhomogeneous microstructure of segregation of elements during the cooling. This study aims to design and fabricate high-entropy alloy with powder metallurgy. In this study, a refractory high entropy alloys with composition near AlMo$_{0.5}$NbTa$_{0.5}$TiZr were produced by powder metallurgy. The alloy consists of two body-centered cubic (BCC) phases. One phase was disordered BCC enriched with Mo, Nb and Ta and the other phase was ordered BCC enriched with Al and Zr. The AlMo$_{0.5}$NbTa$_{0.5}$TiZr alloy had a density of 7.46g/cm$^3$ and Vickers microhardness of 678HV. Its compressive yield strength was 2466MPa at 298K and 964MPa at 1273K. The properties of the alloy and the beneficial effects from powder metallurgy on the microstructure and properties were outlined.

## I. INTRODUCTION

Recently, new alloy systems composed of multi-principal elements, named as high entropy alloys (HEAs), have come to the attention of the scientific community due to the unexpected microstructure and properties. Different from the traditional alloys which are composed of only one principal element with concentration exceeding 50 at. %, HEAs consist of 5 to 13 principal metallic elements with 5~35 at. % for each element. Based on Gibbs Phase Rule (p=n+2-f), the alloy systems that include plenty of elements always show complex phase structure. However, because of the high mixing entropy of HEAs ($\Delta S_{mix}$ is always larger than 1.5R), the systems tend to have negative Gibbs free energy and thus form simple solid solution phases. In HEAs, the effective diffusion rate is limited. These Sluggish Diffusion Effects lead to the appearance of nano-size precipitates. Meanwhile, due to the difference of the atoms in HEAs, the multi-element lattice is highly distorted. The solid solution microstructure, nano-precipitates and distorted lattice together result in the excellent mechanical properties of HEAs [1, 2, 3].

The excellent strength of HEAs, especially at elevated temperature, results in their broad applications in various fields, such as aerospace and nuclear industry. O. N. Senkov et al. (2010) have developed two kinds of refractory HEAs, NbMoTaW and VNbMoTaW, with excellent strength at room temperature and elevated temperature, however, weak ductility and high density [4]. In 2013, they developed AlMo$_{0.5}$NbTa$_{0.5}$TiZr HEA with two BCC solid solution phases and basket-weave lamellar structure [5,6]. It has outstanding room temperature strength, reasonable ductility and low density. However, its strength at the temperature higher than 800°C needs to be enhanced. In the current research, most of the techniques focused on the arc-melting and casting process. To avoid the weaknesses of coarsening of grains and inhomogeneous microstructure of segregation of elements during the cooling in the case of arc-melting and casting process, some HEA systems fabricated by the powder metallurgical process have been reported recently [7,8,9].

In this research, AlMo$_{0.5}$NbTa$_{0.5}$TiZr HEA was fabricated by powder metallurgy. Its microstructure and mechanical properties at both room temperature and elevated temperature are analyzed.

## II. METHODOLOGY

AlMo$_{0.5}$NbTa$_{0.5}$TiZr HEA was fabricated by powder metallurgy. Aluminum, Titanium, Niobium, Molybdenum, and Tantalum were powders in the form of 45μm diameter with purities of 99.9% (KOJUNDO CHEMICAL). Filtration and drying in a vacuum at 120°C were used to extract 150μm-diameter (-100 mesh) Zirconium powder from the stabilizer. Mechanical alloying was carried out in high energy planetary ball mill (Fritsch Pulverisette) at 200 rpm with a ball to powder weight ratio of 15:1 for 6h in Argon atmosphere (Oxygen contents < 80ppm, Humidity < 40ppm). SKD11 vials and Tungsten balls were used as a milling media. The high entropy alloy powders with homogeneous microstructure were compacted at 50 MPa and sintered in the vacuum atmosphere. The HEA was heated to 600°C in 1min and then heated to 1350°C with heating rate as 10°C/min. The sintered HEA was hold at 1350°C for 5min and then cooled in the vacuum chamber. The density of the sintered HEA bulk was measured by Archimedes' principle. Both milled powders and sintered bulk were analyzed by X-ray diffractometer (XRD) with a Cu Kα radiation. The microstructure of sintered HEA bulk was determined by scanning electron microscope (SEM). The chemical composition of each phase in bulks was determined by EPMA. The nanocrystalline nature and the crystal structure of the sintered HEA bulk were analyzed by transmission electron microscope (TEM). The compression properties were measured at both room temperature and 1000˚C.

## III. RESULTS

1. Microstructure analysis

In this research, we fabricated the heat-resistant, high-strength AlMo$_{0.5}$NbTa$_{0.5}$TiZr high-entropy alloy by powder metallurgy. Mechanical alloying was carried out in high energy planetary ball mill at 200rpm with W balls and a ball to powder weight ratio of 15:1 for 6h in Ar atmosphere. Fig.1 shows the crystalline structure of HEA powders before and after mechanical alloying measured by XRD. Two BCC phases can be observed after mechanical alloying.

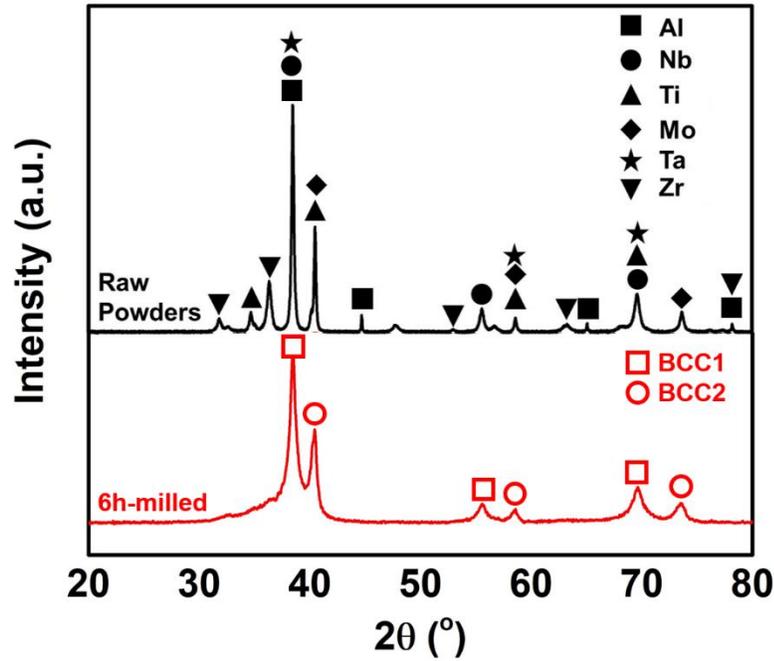

Fig.1 XRD results of AlMo$_{0.5}$NbTa$_{0.5}$TiZr HEA powders before and after mechanical alloying

The HEA powders with homogeneous microstructure were compacted at 50MPa and sintered at 1350°C for 5min. the density of as-sintered HEA bulk is 7.46 g/cm$^3$. The relative density after powder metallurgical process is higher than 99.9%. Two stable BCC phases and Zr oxide phase existed at homogeneous bulk, as shown in the XRD patterns in Fig.2.

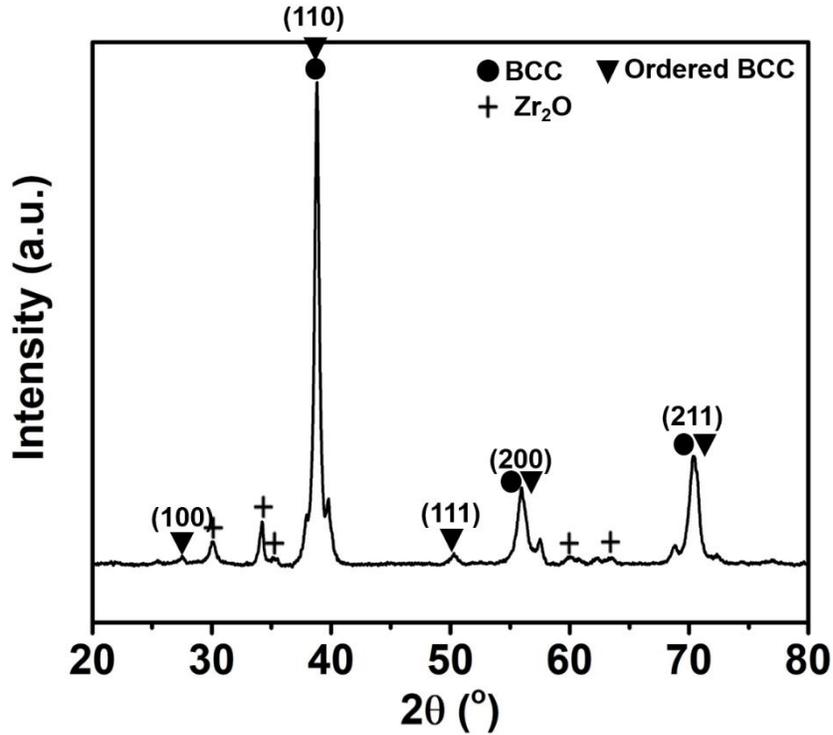

Figure 2. XRD result of as-sintered AlMo$_{0.5}$NbTa$_{0.5}$TiZr HEA

Fig. 3 shows microstructures of as-sintered AlMo$_{0.5}$NbTa$_{0.5}$TiZr HEA. Fig. 4 shows SAD patterns of as-each phase in sintered HEA. 3 phases are observed. Compositions of each phase are listed in Table 1. A (bright matrix) phase is disordered BCC, B (dark grains) phase is ordered BCC (B2), while C (black particles) is Zr oxide phase. Table 2 shows the details of volume fractions and grain sizes of each phase in HEA.

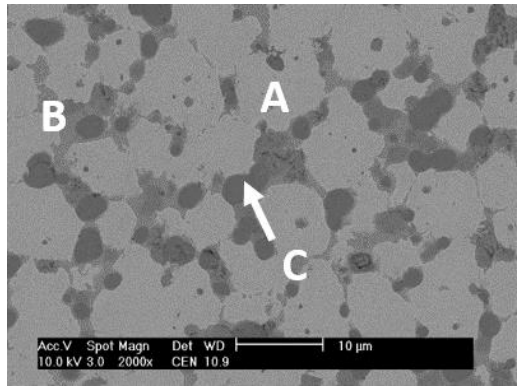

Fig. 3 Microstructures of sintered AlMo$_{0.5}$NbTa$_{0.5}$TiZr HEA bulk

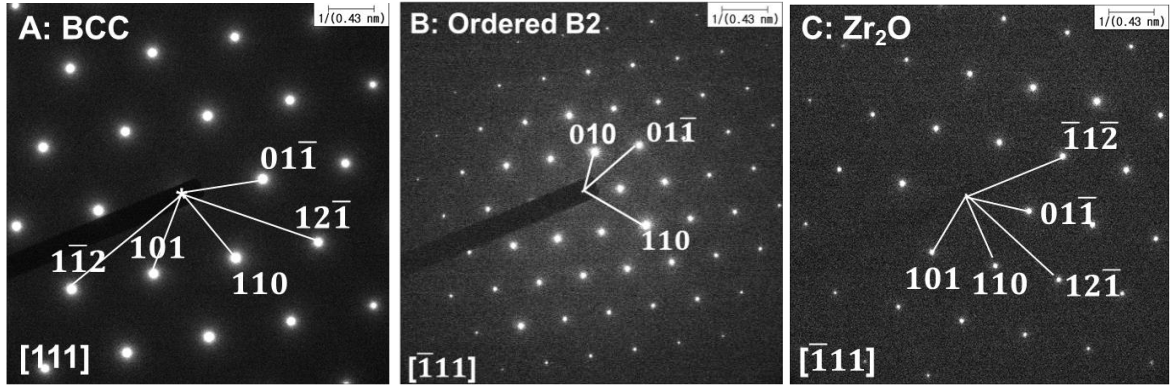

Fig. 4 Diffraction patterns of each phase in as-sintered AlMo$_{0.5}$NbTa$_{0.5}$TiZr HEA measured by TEM

Table 1 Composition of each phase in AlMo$_{0.5}$NbTa$_{0.5}$TiZr HEA

| Region | Al | Mo | Nb | Ta | Ti | Zr | O |
|---|---|---|---|---|---|---|---|
| A: BCC1 | 21.34 | 13.56 | 24.60 | 12.95 | 20.61 | 5.30 | 1.63 |
| B: BCC2 | 30.48 | 4.08 | 11.19 | 8.49 | 17.02 | 25.81 | 2.94 |
| C: Zr$_2$O | 3.098 | 0.904 | 0 | 1.454 | 22.681 | 56.673 | 15.165 |

Table 2 Average grain size and volume fraction of each phase in AlMo$_{0.5}$NbTa$_{0.5}$TiZr HEA

| Region | Grain size (μm) | Volume fraction (%) |
|---|---|---|
| A: BCC1 | 19.8 | 69.7 |
| B: BCC2 | 2.98 | 20.7 |
| C: Zr$_2$O | 2.8 | 6.6 |

2. Mechanical Properties analysis

Microhardness of AlMo$_{0.5}$NbTa$_{0.5}$TiZr HEA is 678HV. Compression properties of AlMo$_{0.5}$NbTa$_{0.5}$TiZr HEA are analyzed at both room temperature and 1000°C. Stress-strain curves are shown in Fig. 5. Yield strength, Young's Modulus, and elongation of HEA at both room temperature and 1000°C are listed in Table 3.

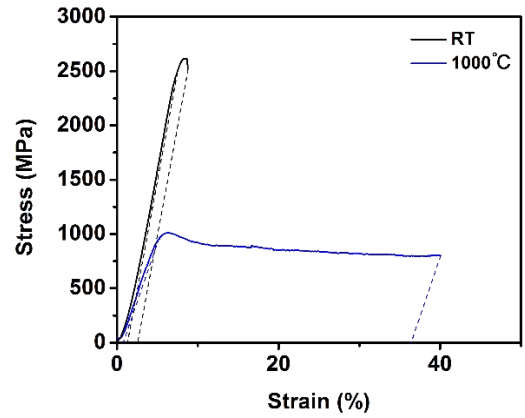

Fig. 5 stress-strain curves of AlMo$_{0.5}$NbTa$_{0.5}$TiZr HEA at room temperature and 1000°C

Table 3 Compression yield strength ($\sigma_{0.2}$), Young's modulus (E), and elongation ($\varepsilon$) of the AlMo$_{0.5}$NbTa$_{0.5}$TiZr alloy at room temperature and 1000°C.

| Test T (°C) | $\sigma_{0.2}$ (MPa) | E (GPa) | $\varepsilon$(%) |
|---|---|---|---|
| Room temperature | 2466 | 40.5 | 8.8 |
| 1000 | 964 | 21.7 | 40.0 |

## IV. DISCUSSION

Compared with arc-melted HEA, the compressive yield strength is improved at both room temperature and 1000°C [5]. Powder metallurgical process provides finer grain size and more homogeneous microstructure. Considering the grain size strengthening mechanism, the finer grain size, the higher strength. That can explain the higher yield strength of as-sintered HEA is higher than that of as-cast HEA at both room temperature and elevated temperature. However, lower plasticity was observed in as-sintered HEA due to more precipitation after powder metallurgy.

Due to the high strength and low density of AlMo$_{0.5}$NbTa$_{0.5}$TiZr HEA, its specific yield strength is excellent compared with refractory HEAs in other literatures, as shown in Fig. 7.

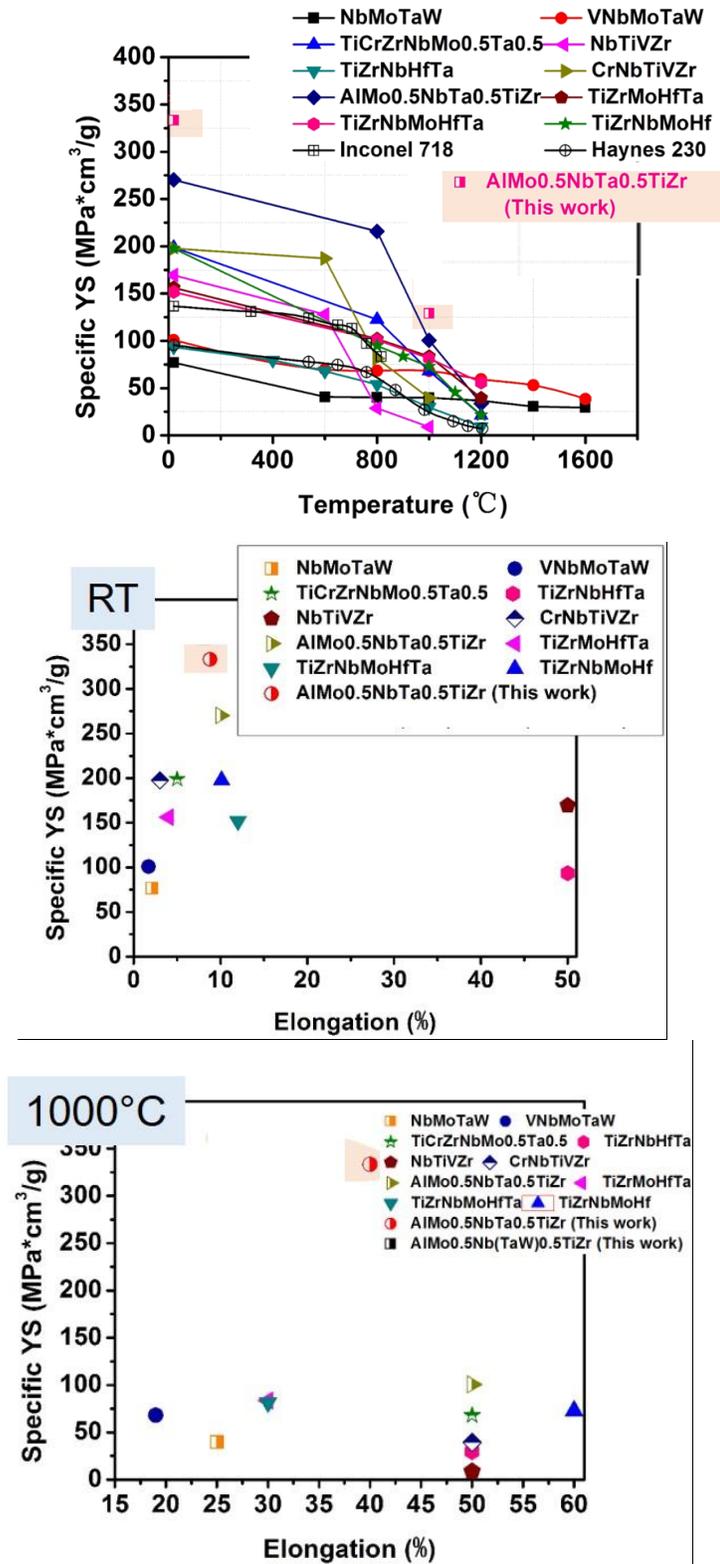

Fig. 7 Comparison of specific yield strength of HEA in this work and previous research at room temperature and elevated temperature

# Effect of W on microstructures and elevated-temperature mechanical properties of refractory AlMo$_{0.5}$Nb(TaW)$_{0.5}$TiZr High Entropy Alloy

Yuqiao Li, Junho Lee, Byungchul Kang, Soon Hyung Hong

Department of Materials Science and Engineering

Korea Advanced Institute of Science and Technology, Republic of Korea

**Abstract**

New approaches for the design of alloy systems with multiprincipal elements is recently researched in refractory materials field. However, no HEA with excellent specific strength at temperature higher than 800°C was reported. This study aims to design and fabricate high-entropy alloy with low density and high specific strength at elevated temperature by powder metallurgy. In this study, a refractory high entropy alloys with composition near AlMo$_{0.5}$Nb(TaW)$_{0.5}$TiZr were produced by powder metallurgy. The alloy consists of two body-centered cubic (BCC) phases. One phase was disordered BCC enriched with Mo, Nb, W and Ta and the other phase was ordered BCC enriched with Al and Zr. The AlMo$_{0.5}$Nb(TaW)$_{0.5}$TiZr alloy had a density of 7.56 g/cm$^3$ and Vickers microhardness of 781 HV. Its yield strength was 2684 MPa at 298 K and 1080 MPa at 1273 K. The properties of the alloy and the beneficial effects from powder metallurgy on the microstructure and properties were outlined.

## VI. INTRODUCTION

Recently, new alloy systems composed of multi-principal elements, named as high entropy alloys (HEAs), have come to the attention of the scientific community due to the unexpected microstructure and properties. Different from the traditional alloys which are composed of only one principal element with concentration exceeding 50 at. %, HEAs consist of 5 to 13 principal metallic elements with 5~35 at. % for each element. Based on Gibbs Phase Rule (p=n+2-f), the alloy systems that include plenty of elements always show complex phase structure. However, because of the high mixing entropy of HEAs ($\Delta S_{mix}$ is always larger than 1.5R), the systems tend to have negative Gibbs free energy and thus form simple solid solution phases. In HEAs, the effective diffusion rate is limited. These Sluggish Diffusion Effects lead to the appearance of nano-size precipitates. Meanwhile, due to the difference of the atoms in HEAs, the multi-element lattice is highly distorted. The solid solution microstructure, nano-precipitates and distorted lattice together result in the excellent mechanical properties of HEAs [1, 2,3].

The excellent strength of HEAs, especially at elevated temperature, results in their broad applications in various fields, such as aerospace and nuclear industry. O. N. Senkov et al. (2010) have developed two kinds of refractory HEAs, NbMoTaW and VNbMoTaW, with excellent strength at room temperature and elevated temperature, however, weak ductility and high density [4]. In 2013, they developed AlMo$_{0.5}$NbTa$_{0.5}$TiZr HEA with two BCC solid solution phases and basket-weave lamellar structure [5,6]. It has outstanding room temperature strength, reasonable ductility and low density. However, its strength at the temperature higher than 800°C needs to be enhanced. With high melting temperature, Tungsten is widely used in heat-resistant High Entropy Alloys.

In this research, AlMo$_{0.5}$Nb(TaW)$_{0.5}$TiZr HEA was fabricated by powder metallurgy for more homogeneous microstructure and better mechanical properties. Its microstructure and mechanical properties at both room temperature and elevated temperature are analyzed.

## VII. METHODOLOGY

AlMo$_{0.5}$NbTa$_{0.5}$TiZr HEA was fabricated by powder metallurgy. Aluminum, Titanium, Niobium, Molybdenum, and Tantalum were powders in the form of 45µm diameter with purities of 99.9% (KOJUNDO CHEMICAL). Filtration and drying in a vacuum at 120°C were used to extract 150µm-diameter (-100 mesh) Zirconium powder from the stabilizer. Mechanical alloying was carried out in high energy planetary ball mill (Fritsch Pulverisette) at 200 rpm with a ball to powder weight ratio of 15:1 for 6h in Argon atmosphere (Oxygen contents < 80ppm, Humidity < 40ppm). SKD11 vials and Tungsten balls were used as a milling media. The high entropy alloy powders with homogeneous microstructure were compacted at 50 MPa and sintered at 1400 °C for 5 min in the vacuum atmosphere. The density of the sintered HEA bulk was measured by Archimedes' principle. Both milled powders and sintered bulk were analyzed by X-ray diffractometer (XRD) with a Cu Kα radiation. The microstructure of sintered HEA bulk was determined by scanning electron microscope (SEM). The chemical composition of each phase in bulks was determined by EPMA. The nanocrystalline nature and the crystal structure of the sintered HEA bulk were analyzed by transmission electron microscope (TEM). The mechanical properties including compression and tension were measured at both room temperature and 1000˚C.

## VIII. RESULTS

3.  Microstructure analysis

In this research, we fabricated the heat-resistant, high-strength AlMo$_{0.5}$Nb(TaW)$_{0.5}$TiZr high-entropy alloy by powder metallurgy. Mechanical alloying was carried out in high energy planetary ball mill at 200rpm with W balls and a ball to powder weight ratio of 15:1 for 6h in Ar atmosphere. Fig.1 shows the crystalline structure of HEA powders before and after mechanical alloying measured by XRD. Two BCC phases can be observed after mechanical alloying, while ZrH$_2$ also shows its peaks.

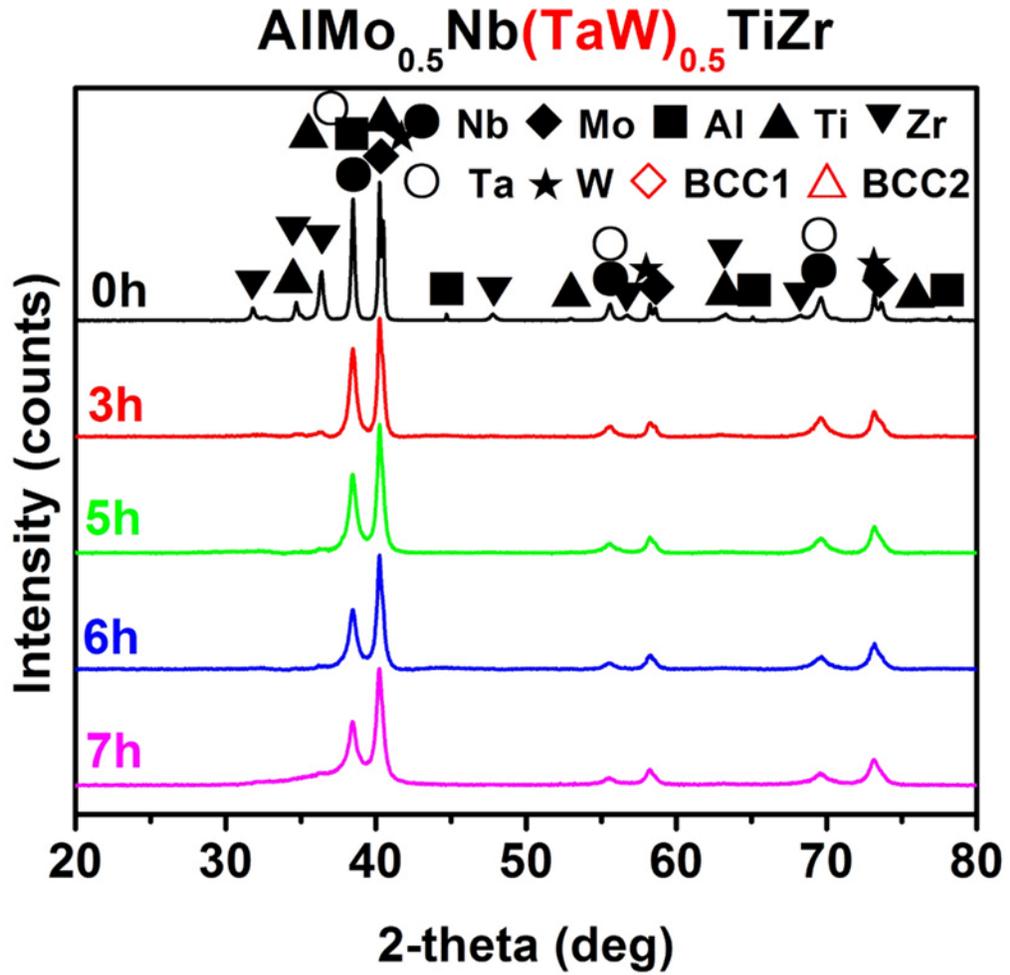

Fig.1 XRD results of HEA powders before and after mechanical alloying measured

The HEA powders with homogeneous microstructure were compacted at 50MPa and sintered at 1400°C for 5min. Two stable BCC phases existed at homogeneous powders and bulks as shown in the XRD patterns in Fig.2.

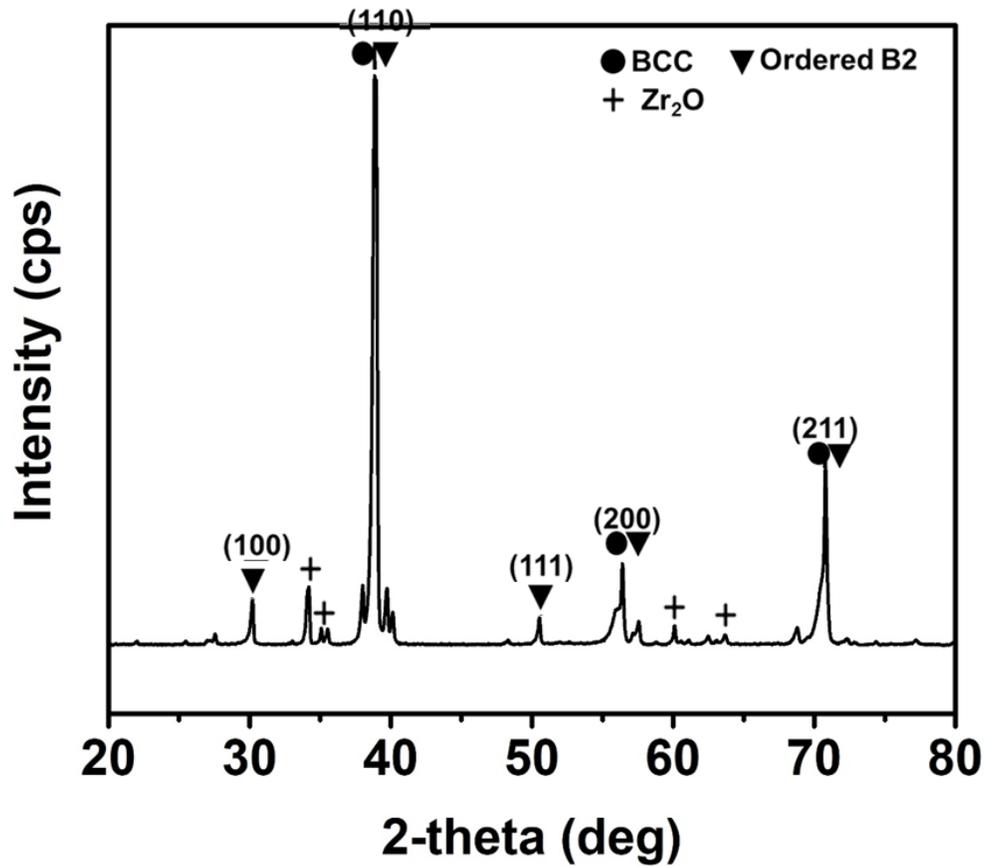

Figure 2. XRD result of AlMo$_{0.5}$NbTa$_{0.5}$TiZr HEA bulk after sintering.

Fig. 3 shows microstructures of sintered HEA. It's composed of 3 phases. Fig. 4 shows TEM microstructure and SAD patterns of 1400°C-sintered HEA. 3 phases are observed. Compositions of each phase are listed in Table 1. A (bright matrix) phase is disordered BCC enriched with Mo, Nb, Ta and W, B (dark grains) phase is ordered B2 enrich with Al and Zr, while C (black particles) is Zr oxide phase. Table 2 shows the details of volume fractions and grain sizes of each phase in HEA.

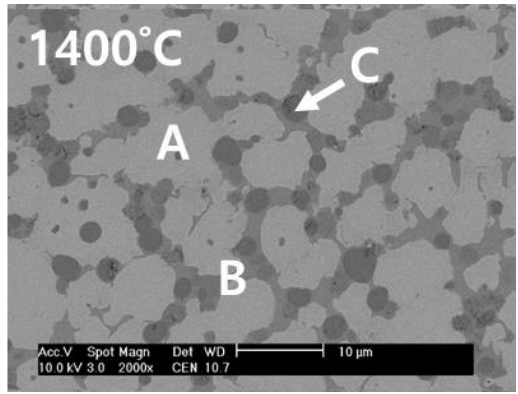

Fig. 3 Microstructures of sintered AlMo$_{0.5}$NbTa$_{0.5}$TiZr HEA bulk

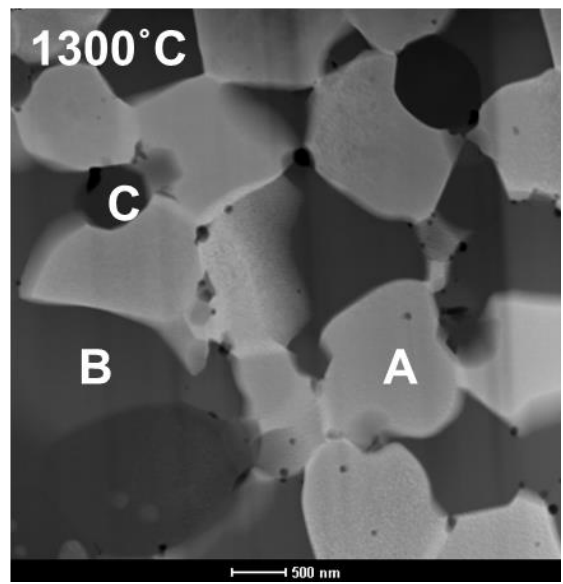

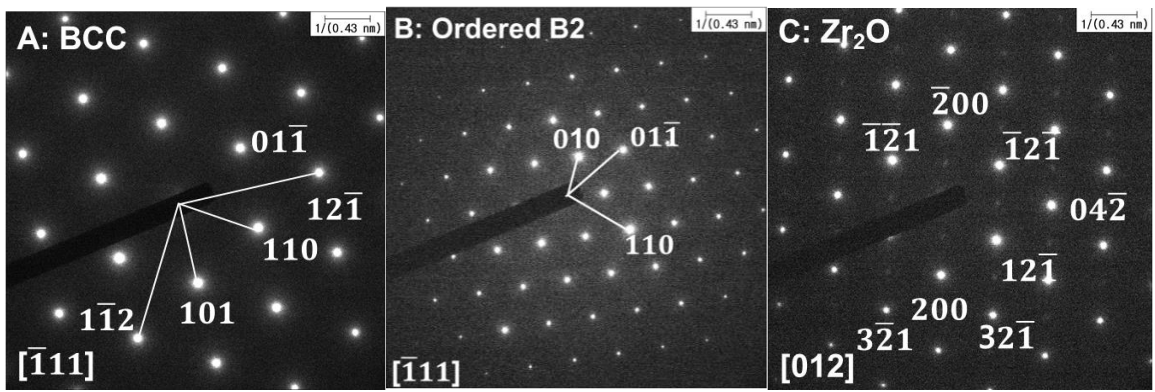

Fig. 4 Microstructures and diffraction patterns of each phase in sintered AlMo$_{0.5}$NbTa$_{0.5}$TiZr HEA measured by TEM

Table 1 Composition of each phase in AlMo$_{0.5}$NbTa$_{0.5}$TiZr HEA

| Region | Al | Mo | Nb | Ta | Ti | Zr | W | O |
|---|---|---|---|---|---|---|---|---|
| A: Disordered BCC | 20.63 | 15.00 | 26.94 | 7.89 | 19.73 | 4.52 | 5.29 | - |
| B: Ordered B2 | 41.84 | 4.18 | 5.36 | 1.84 | 14.36 | 32.09 | 0.32 | - |
| C: Zr$_2$O | 2.06 | 0.64 | 0.00 | 0.36 | 10.32 | 26.47 | 0.18 | 59.98 |

Table 2 Average grain size and volume fraction of each phase in AlMo$_{0.5}$Nb(TaW)$_{0.5}$TiZr HEA

| Grain size of BCC1 phases (μm) | Volume fraction of BCC1 (%) | Grain size of BCC2 phases (μm) | Volume fraction of BCC2 (%) | Grain size of Zr$_2$O (μm) | Volume fraction of Zr$_2$O (%) |
|---|---|---|---|---|---|
| 3.10 | 70.824 | 2.98 | 20.669 | 0.182 | 8.507 |

4. Mechanical Properties analysis

Micro hardness of AlMo$_{0.5}$NbTa$_{0.5}$TiZr HEA is 781HV. Compression properties of AlMo$_{0.5}$NbTa$_{0.5}$TiZr HEA are analyzed at both room temperature and 1000°C. Stress-strain curves are shown in Fig. 5. Yield strength, Young's Modulus, and elongation of HEA at both room temperature and 1000°C are listed in Table 3.

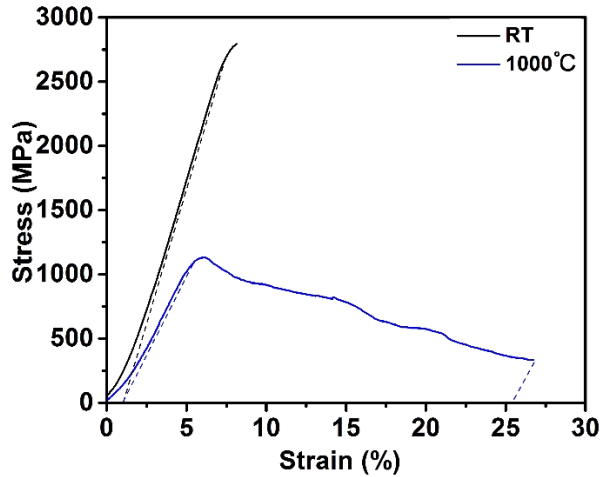

Fig. 5 Stress-strain curves of AlMo$_{0.5}$NbTa$_{0.5}$TiZr HEA at room temperature and 1000°C

Table 3 Compression yield strength ($\sigma_{0.2}$), Young's modulus (E), and elongation ($\varepsilon$) of the AlMo$_{0.5}$Nb(TaW)$_{0.5}$TiZr alloy at room temperature and 1000°C.

| Test T (°C) | $\sigma_{0.2}$ (MPa) | E (GPa) | $\varepsilon$(%) |
|---|---|---|---|
| Room temperature | 2684 | 41.7 | 8.14 |
| 1000 | 1080 | 24.3 | 26.7 |

## IX. DISCUSSION

Compared with AlMo$_{0.5}$NbTa$_{0.5}$TiZr HEA in another research, AlMo$_{0.5}$Nb(TaW)$_{0.5}$TiZr HEA has a higher yield strength at both room temperature and 1000 °C, with relatively reduced elongation and plastic deformation. Meanwhile, its grain size is decreased and its precipitation volume fraction is increased, as shown in Table 4.

Table 4 Comparison between AlMo$_{0.5}$NbTa$_{0.5}$TiZr HEA and AlMo$_{0.5}$Nb(TaW)$_{0.5}$TiZr HEA, where G.S. is average grain size, V.f.of prec.is volume fraction of precipitation, $\sigma_{0.2}$ is yield strength, $\sigma_m$ is maximum compression strength, E is Young's Modulus, $\varepsilon_f$ is fracture elongation, $\varepsilon_p$ is plastic deformation.

| Alloys | G.S. (μm) | V.f.of prec. (%) | Testing T (°C) | $\sigma_{0.2}$ (MPa) | $\sigma_m$ (MPa) | E (GPa) | $\varepsilon_f$(%) | $\varepsilon_p$(%) |
|---|---|---|---|---|---|---|---|---|
| AlMo$_{0.5}$NbTa$_{0.5}$TiZr | 13.4 | 30.3 | RT | 2486 | 2617 | 40.5 | 8.76 | 1.3 |
| | | | 1000 | 964 | 1012 | 21.7 | 40.0 | 35.6 |
| AlMo$_{0.5}$Nb(TaW)$_{0.5}$TiZr | 6.7 | 34.1 | RT | 2684 | 2794 | 41.7 | 8.14 | 0.4 |
| | | | 1000 | 1080 | 1132 | 24.3 | 26.7 | 24.4 |

With the addition of W, the atomic difference of HEA [3] is increased from 3.24 to 3.45, which means there is more severe lattice distortion in the HEA with W. With more dislocations, the grains size of HEA is decreased. Based on Hall-Petch Equation, the strength is increased with lower grain size. The Hall-Petch relationship of HEAs with and without W is shown in Fig. 6. On the other hand, more severe lattice distortion hinders atomic movement which limits diffusion rate,and thus causes more precipitation. In AlMo$_{0.5}$Nb(TaW)$_{0.5}$TiZr HEA the precipitation is Zr Oxide, which is more brittle than BCC matrix. The ductility is decreased with more precipitation. Fig 7 shows the DSC curve of HEAs with and without W. The endothermic line shows the collapse of crystalline structure of HEAs under high temperature. The exothermic line under 500 °C shows the release of internal stresses, such as structural deformation, lattice strain, etc. Shorter exothermic region of HEA with W means there is less internal stress in the HEA and causes higher strength at room temperature compared HEA without W. After the release of internal stresses, there are continues endothermic lines for both HEAs. At 1000 ˚C, HEA without W has finished the collapse of crystalline structure and started phase transformation, while HEA with W has not, which causes the higher strength at 1000 ˚C. The exothermic peaks at 1270 ˚C shows a release of energy due to phase transformation process.

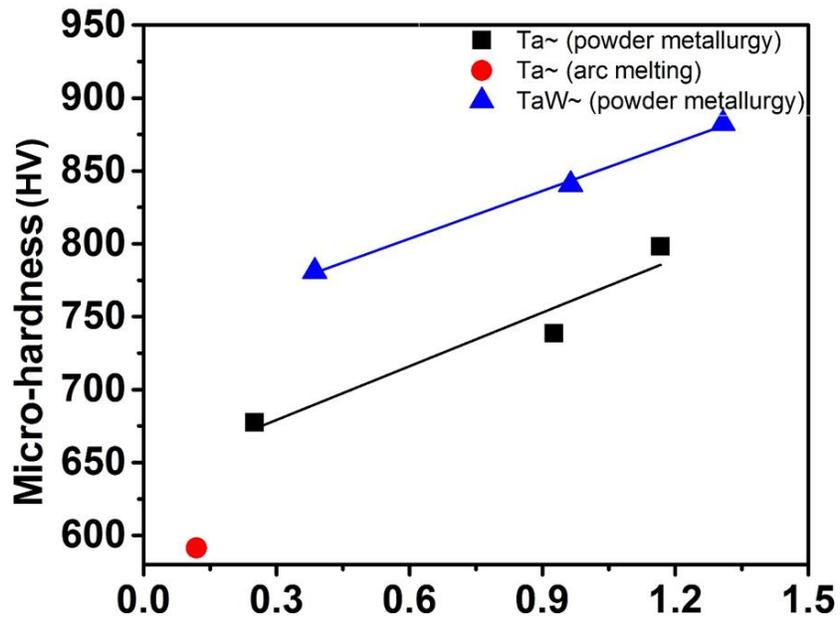

Fig. 6 Hall-Petch relationship between HEAs with and without W.

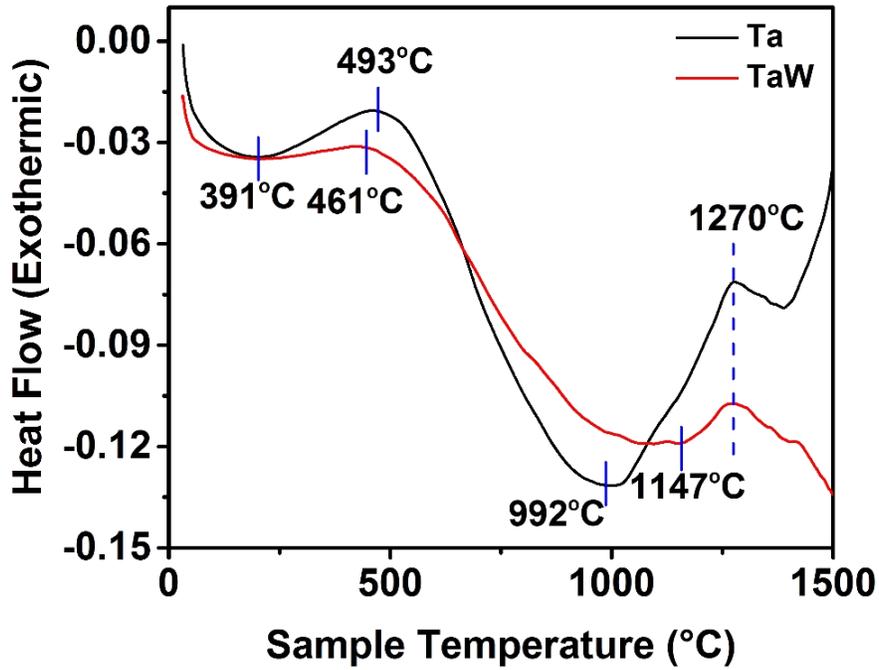

Fig. 7 DSC curves of HEAs with and without W

X. CONCLUSION

With the addition of W, AlMo$_{0.5}$Nb(TaW)$_{0.5}$TiZr HEA has an increased strength at both room temperature and high temperature, with a relatively reduced ductility. Compared with other refractory HEAs, AlMo$_{0.5}$Nb(TaW)$_{0.5}$TiZr HEA shows more excellent specific mechanical properties, as shown in Fig.8.

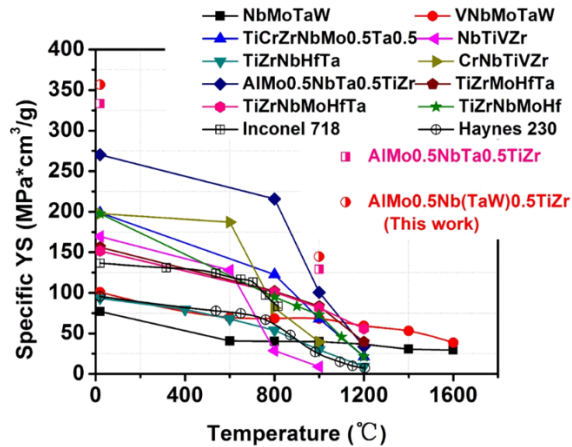

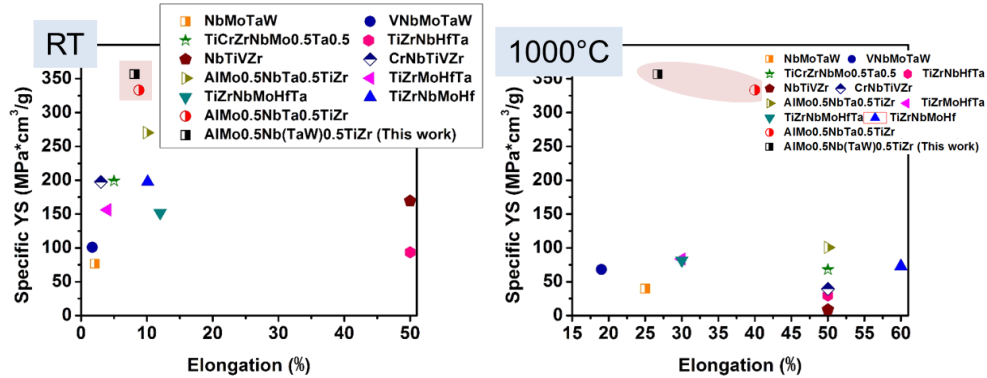

Fig. 9 Comparison among HEAs. AlMo$_{0.5}$Nb(TaW)$_{0.5}$TiZr HEA shows more excellent mechanical properties.